\documentclass[aps,prr,superscriptaddress,twocolumn,showpacs]{revtex4-1}

\usepackage[table]{xcolor}
\usepackage{pgf}
\usepackage[utf8]{inputenc}
\usepackage{amsmath}
\usepackage{braket}
\usepackage{amssymb}
\usepackage{amsmath}
\usepackage{bbold}
\usepackage{enumitem} 
\usepackage{appendix}
\usepackage{graphicx}
\usepackage{overpic}
\usepackage{epstopdf}
\usepackage[pdftex,bookmarks,colorlinks]{hyperref}
\usepackage{hyperref}

\usepackage{ulem}

\newcommand{\eq}[1]{Eq.~(\ref{#1})}

\definecolor{caribbeangreen}{rgb}{0.0, 0.8, 0.6}

\newcommand{\proj}[1]{\ket{#1}\!\!\bra{#1}}
\newcommand{\mhz}{\,\mathrm{MHz}}
\newcommand{\ghz}{\,\mathrm{GHz}}

\begin{document}
\title{Real-time-dynamics quantum simulation of (1+1)-D lattice QED with Rydberg atoms}

\author{Simone Notarnicola}
\affiliation{Dipartimento di Fisica e Astronomia ``G. Galilei'', via Marzolo 8, I-35131, Padova, Italy}
\affiliation{INFN, Sezione di Padova, via Marzolo 8, I-35131, Padova, Italy}

\author{Mario Collura}
\affiliation{Dipartimento di Fisica e Astronomia ``G. Galilei'', via Marzolo 8, I-35131, Padova, Italy}
\affiliation{SISSA , Via Bonomea 265, I-34136 Trieste, Italy}

\author{Simone Montangero}
\affiliation{Dipartimento di Fisica e Astronomia ``G. Galilei'', via Marzolo 8, I-35131, Padova, Italy}
\affiliation{INFN, Sezione di Padova, via Marzolo 8, I-35131, Padova, Italy}

\begin{abstract}
We show how to implement a Rydberg-atom quantum simulator to study the non-equilibrium dynamics of an Abelian (1+1)-D  lattice gauge theory.
The implementation locally codifies the degrees of freedom of a $\mathbf{Z}_3$  gauge field,
once the matter field  is integrated out by means of the Gauss' local symmetries. 
The quantum simulator scheme is based on current available technology and scalable to considerable lattice sizes. It allows, within experimentally reachable regimes,  
to explore different string dynamics and to infer information about the Schwinger $U(1)$ model. 

\end{abstract}
\maketitle

%
{
Rydberg-atom systems are nowadays one of the most promising and versatile platform in the field of quantum simulation for the achievement of results unaccessible via classical numerical simulations~\cite{glaetzle2014quantum, Labuhn2016, Bernien2017,PhysRevX.8.021070,Weimer2010, Bernien2017, omran2019generation,wintermantel2019unitary}.
The internal ground state of neutral atoms is coupled to an highly excited Rydberg state, realizing a benchmarked qubit prototype~\cite{PhysRevA.97.053803,buluta2011natural,Gaetan2009,PhysRevLett.121.123603}.
The strong dipole-dipole coupling between excited atoms induces the Rydberg blockade mechanism~\cite{PhysRevLett.112.183002}, allowing to engineer local dynamical constraints. 
Optical tweezer arrays allow to trap and arrange large number of atoms in various geometries, from one dimensional lattices to {unconventional} three-dimensional structures~\cite{Barredo2018,Labuhn2016}.
%
%
Recent experimental results concerning a Rydberg-atom chain have raised the interest in studying possible connections between Abelian lattice gauge theories and Rydberg-atom systems~\cite{Labuhn2016, surace2019lattice}.
Despite the proposals realized in the last decade for studying  Abelian and non-Abelian  lattice gauge theories via universal quantum simulators~\cite{preskill2018simulating,wiese2013ultracold,doi:10.1080/00107514.2016.1151199,
banerjee2012atomic,stannigel2013constrained,wiese2013ultracold,doi:10.1080/00107514.2016.1151199,zohar2012simulating, zohar2013simulating,zohar2013cold,Zache_2018,Tagliacozzo2013,PhysRevLett.111.115303,PhysRevD.95.094507,PhysRevA.94.063641,PhysRevA.100.013629}, 
the search for a mapping suitable for systems with large lattice sizes is still open.
%
%
In recent experiments, trapped ions~\cite{Martinez2016}  and ultracold atoms ~\cite{alex2019realizing,schweizer2019floquet} have been used to explore the mechanism of pairs of particle-antiparticle production in the Schwinger model for minimal-size systems, while hybrid protocols combining quantum simulation~\cite{Kokail2019} or computation~\cite{PhysRevA.98.032331} with classical numerical simulations have been set up. 
The realization of a quantum simulator scalable to large lattice sizes, even for minimal gauge models as the Abelian ones, would path the way to the study of high-energy physics non-local phenomena, such as scattering processes and string breaking \cite{PhysRevD.100.036009,PhysRevX.6.041040}, giving motivation to our work.
%
%
%
}
\begin{figure}
    \centering
    \begin{tabular}{c}
    \includegraphics[width=7.5cm]{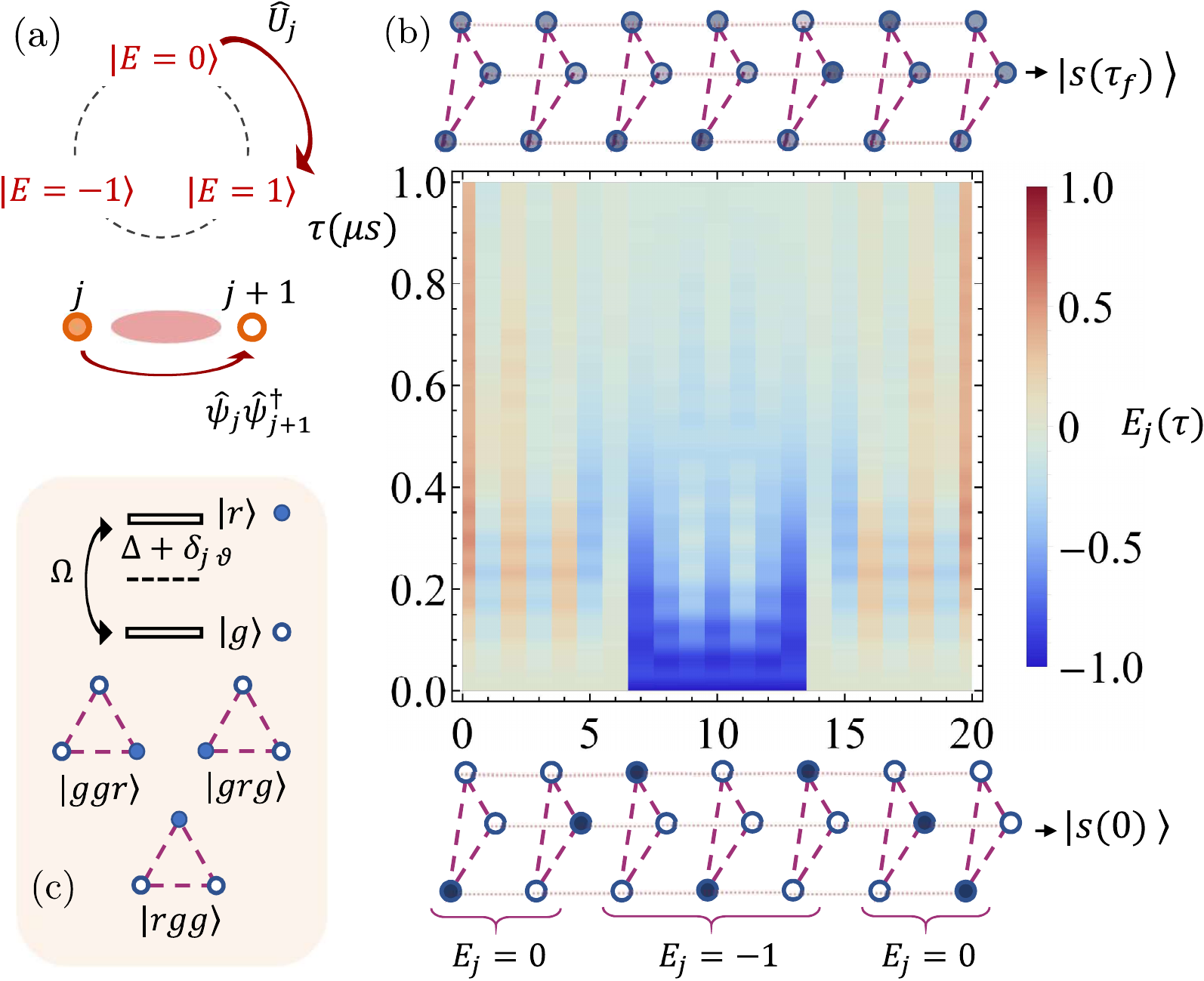}
\end{tabular}
 \caption{\label{fig:intro}
 (a) Dynamics of the $\mathbf{Z}_3$ gauge and fermion field relative to the Hamiltonian $\hat H$ of \eq{eq:Z3_ham_g}: 
 A rotation of the local gauge field state is accompanied by  a fermion hopping process which preserves gauge invariance.
 Given a set of three equidistant atoms, the electric field eigenstates $\ket{E}_j$  are mapped into those configurations with only one excited atom.
(b) Chain of atomic triangular sets. 
An electric field string is represented by the separable state $\ket{s(0)}$ at $\tau=0$: 
By setting the parameters in the Hamiltonian $\hat H_r$ (see \eq{eq:ryd_ham}) the system evolves to the state $\ket{s(\tau_f)}$ in which the initial string is broken.
(c) The ground state of the atoms (empty dot) is coupled  to a Rydberg state (full dot) by a Rabi frequency $\Omega$ with a local blue-shifted detuning $\Delta + \delta_{j \vartheta}$.
Numerical parameters of Hamiltonian in \eq{eq:Z3_ham_g}: $m=0,\,t=0.682 \mhz,\,g^2/t=0.5$. }
  \end{figure} 

{
We propose a Rydberg-atom quantum simulator to study a 1D Abelian lattice gauge theory with spinless fermions 
coupled to an electric field. Fermions live on the lattice sites while the electric field is defined on each link between two neighbouring sites.
In particular, we map the dynamics of our simulator into the gauge invariant dynamics of the electric field. We focus on an electric field string generated by two opposite charges put apart on the lattice to study different dynamical regimes: the string can persist in time or can be broken by the spontaneous creation of particle-antiparticles pairs in the middle of it, in analogy with the confinement properties of   the QCD~\cite{PhysRevX.6.011023, PhysRevLett.109.175302, PhysRevLett.102.191601}.
}

{
Encoding a quantum gauge field into atomic degrees of freedom  imposes to discretize and truncate its spectrum: Starting from the Schwinger model, we replace the continuous symmetry group $U(1)$ with $\mathbf{Z}_n$ in which the electric field spectrum contains $n$ discrete values.
We choose $n=3$ so that the spectrum of the local electric field is ($0,\,\pm1$) -- Fig.~\ref{fig:intro} (a).
 %
%
%
Differently from the well-studied case $n=2$\cite{surace2019lattice,celi2019emerging}, in which the spectrum is $(\pm 1/2)$, in our case the electric field energy contribution is non trivial (see \eq{eq:Z3_ham_g}): We can therefore induce different string behaviors by changing the values of the couplings.  
The $\mathbf{Z}_3$ gauge symmetry is encoded in the lattice geometry of our quantum simulator, as shown in Fig.~\ref{fig:intro} (b): Groups of three Rydberg atoms represent each link, and we map the eigenstates of the local electric field into the configurations with only one atom excited -- Fig.~\ref{fig:intro} (c). 
By applying an effective Rabi frequency $\Omega$, we allow each atom to oscillate between the ground state and the Rydberg state.
By tuning their inter-atomic distances, we exploit the Rydberg blockade to suppress those atomic configurations corresponding to states which break the Gauss' law in a given charge sector. 
As a result, we find that the dynamics of the atomic excitations reproduces the gauge invariant dynamics of the $\mathbf{Z}_3$ gauge field.
The Hamiltonian parameters can be modulated by changing a local detuning applied to each atom -- Fig.~\ref{fig:intro} (c).
%
%
 We check the reliability of our quantum simulator in two steps: 
 First, we compare the $\mathbf{Z}_3$ and the $U(1)$ models, discussing under which conditions their dynamics coincide.
 Despite some differences, we find that the parameters regime wherein string-breaking dynamics occurs is the same for both theories.
Second, we compare the dynamics of the $\mathbf{Z}_3$ model with that of our quantum simulator, 
showing that our scheme allows to investigate the two different string dynamics regimes. 
We conclude that our quantum simulator is able to capture reliable features of continuous and discrete Abelian lattice gauge theories.
}

\textit{The model.--- } The Hamiltonian for the (1+1)-D lattice QED in the Schwinger formulation is~\cite{kogut1975hamiltonian,Montvay:1994cy}
\begin{align}\label{eq:QED_ham}
\hat H_{S}=\sum_j\left[-t\,\hat \psi^\dagger_j \mathcal{\hat U}_{j}\hat \psi_{j+1}+\mathrm{h.c.} +m\, p_j\hat n_j
+g^2 \mathcal{\hat E}_{j}^2\right],
\end{align}
where a staggered, fermionic, spinless matter field with mass $m$  is defined on the lattice sites. It satisfies $\{\hat \psi^\dagger_{j},\hat \psi_{k}\}=\delta_{jk}$ and $\{\hat \psi_{j},\hat \psi_{k}\}=\{\hat \psi^\dagger_{j},\hat \psi^\dagger_{k}\}=0$, while $\hat n_j=\hat\psi^\dagger_j \hat\psi_j$ and  $p_x=(-1)^j$. 
The gauge field propagator $\mathcal{\hat U}_{j}$ and the electric field $\mathcal{\hat E}_{j}$ are defined on each link between the nearest-neighbor sites $j,\,j+1$: 
They commute according to $[\mathcal{\hat E}_{j},\mathcal{\hat U}_{k}]=\delta_{jk}\,\mathcal{\hat U}_{j}$ and $g^2$ is the electric field energy coupling. 

Due to the staggering, the electrons (positrons) are represented by filled (empty) even (odd) sites and, therefore, the gauge-matter interaction term proportional to $t$ is responsible for electron-positron pair creation and annihilation. 
During these processes, the electric field is incremented or decremented in order to satisfy the Gauss' law on each site:  
 Equivalently,  given the set of gauge operators $\hat Q_j= \Delta  \mathcal{\hat E}_j-\frac{1-p_j}{2}+e\,\hat n_j$, with $\Delta  \mathcal{\hat E}_j = \mathcal{\hat E}_j-\mathcal{\hat E}_{j-1}$ and $e=-1$, we have $[\hat H_{S},\hat Q_j ]=0\,\forall\,j$. 
 It follows that the Hamiltonian has a block-diagonal form in the basis of the gauge operators' eigenstates: 
 Each block is identified by a set of static charges $g_j$ and the relative states satisfy $\hat Q_j \ket{\psi}=g_j \ket{\psi}\,\forall j$.  

Once the boundary conditions and the set of static charges $g_j$ are fixed, the gauge operators $\hat Q_j$ fix a one-to-one correspondence between the eigenstates of the matter and the electric fields operators $\hat \psi_j^\dagger\hat \psi_j$ and $\mathcal{\hat E}_{j}$, indicated as $\{\ket m\}$ and $\{\ket{\mathcal{E}}\}$ respectively: 
Therefore, the basis of the gauge sector characterized by $\set{g_j}$ is in the form $\ket{m,\mathcal{ E}; \set{g_j}}$. 
It follows that  $\hat H_{S}$ can be recast in each sector  as a function of the matter or the gauge field operators~\cite{PhysRevD.99.114511, PhysRevB.98.075119}. 
Hereafter, we set $g_j=0\,\forall j$ and write the Hamiltonian as a function of the gauge field operators: 
The Hamiltonian  is still local, namely (see the Appendix \ref{app:Ham}) 
\begin{align}\label{eq:QED_ham_g}
\hat H_{S}^g =&\, -t\sum_{j}\mathcal{\hat  U}_j^\dagger\, \hat P_{\Delta \mathcal{E}_{j}}(0) \hat P_{\Delta \mathcal{ E}_{j+1}}(1) + \mathrm{h.c.}  \\ \nonumber
&\,+ m \sum_j \left( \frac{1-p_j}{2}-\Delta\mathcal{\hat E}_j\right)+g^2\sum_j \mathcal{\hat E}^2_j \,.
\end{align}
The projectors  $\hat P_{\Delta \mathcal{E}_{j}}(n_j)$ select the electric field configurations whose expectation values satisfy the local gauge invariance condition $\Delta \mathcal{E}_{j}=\frac{1-p_j}{2}+n_j$.
As a consequence, the hopping matrix elements of the Hamiltonian $\hat H_{S}$ between the states $\ket{m,\mathcal{ E}},\,\ket{m',\mathcal{ E'}}$ coincide with those of  $\hat H^g_{S}$ computed over the states $\ket{\mathcal{ E}},\,\ket{\mathcal{ E'}}$.

    \begin{figure}
    \centering
    \begin{tabular}{c}
    \includegraphics[width=6.5cm]{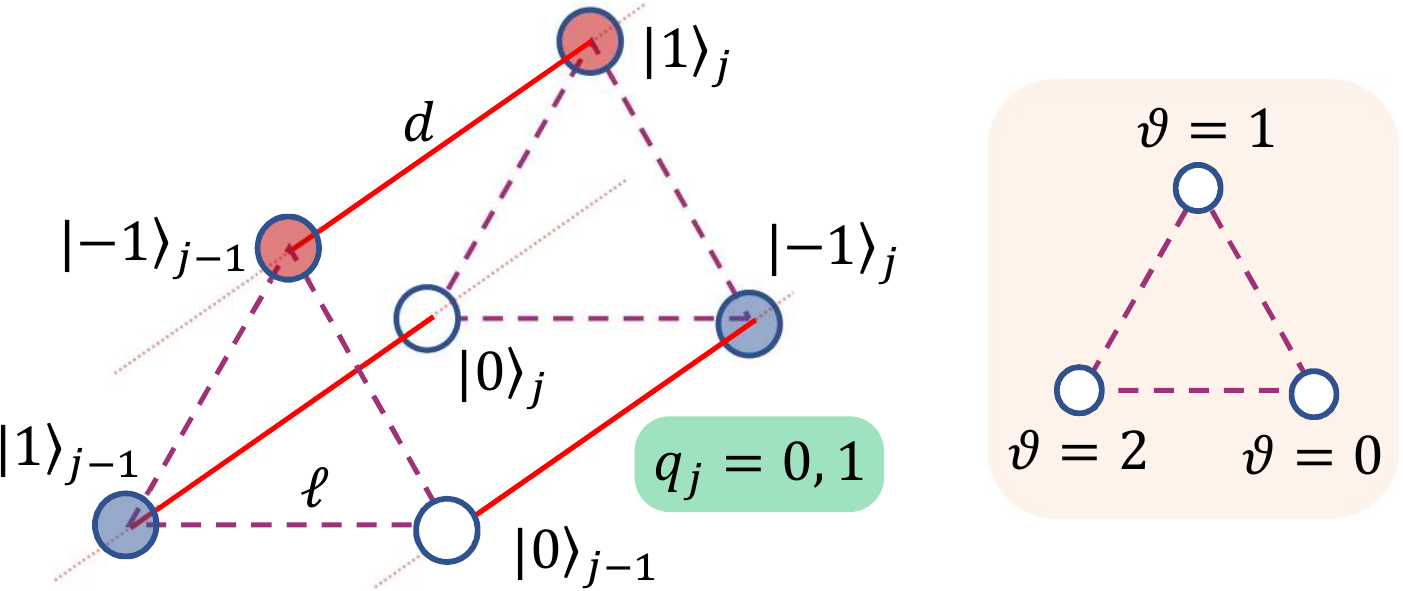} \\ 
     \pgfputat{\pgfxy(-3.5,3.1)}{\pgfbox[left,top]{\footnotesize (a)}}
    \pgfputat{\pgfxy(1.2,3.1)}{\pgfbox[left,top]{\footnotesize (b)}}      
     \end{tabular}
 \caption{\label{fig:shape}Atomic lattice geometry:  (a)
 The distances $\ell$ and $d$  are short enough such that Rydberg blockade prevents simultaneous excitations of atoms in the same link or in different links but aligned along the lattice axes (filled red circles). 
Simultaneous excitations of non aligned atoms from different links are  allowed (blue filled circles).  
(b) A label $\vartheta$ is assigned to the atoms, uniformly for each link, to map each $\ket{E}_j$ into the states $\ket{\vartheta}_j$.
}
  \end{figure} 

In order to encode the gauge degrees of freedom in a quantum simulator,  we need to truncate and discretize the spectrum of the electric field.
To this purpose, we replace the continuous-spectrum operator $\mathcal{\hat  U}_j$ with the discrete clock operator ${\hat  U}_j$ such that ${\hat  U}_j^n=(\hat  U^\dagger_j)^n=\mathbf{1}$ with $n \in \mathbb{N}$. 
That is, we move from the continuous gauge symmetry group $U(1)$ to  $\mathbf{Z}_n$~\cite{Notarnicola_2015,PhysRevD.98.074503}. 
We  fix $n=3$, so that the electric field $\hat E_{j}$ admits only three possible states 
$\{\ket{- 1}, \, \ket{0},\,\ket{ + 1} \}$
and the operators ${\hat  U}_j\,({\hat  U}_j^\dagger)$  cyclically permute them clockwise (anti-clockwise) as shown in FIG~\ref{fig:intro} (a). 
Gauge invariance is guaranteed by the condition 
$\tilde\Delta \hat{E}_{j}= (\frac{1-p_j}{2}+\hat n_j)  \mod 3$ with spectrum  $\{-1,0,1\}$.
Finally, we define the electric field energy to be proportional to $\sum_j\hat E^2_j$. In conlusion, the Hamiltonian reduces to
\begin{align}\label{eq:Z3_ham_g}
\hat H=&\, -t\sum_{j} \hat U_j^\dagger\, \hat P_{\tilde\Delta E_{j}}(0) \hat P_{\tilde\Delta E_{j+1}}(1) + \mathrm{h.c.}  \\ \nonumber
&\,+ m \sum_j \left( \frac{1-p_j}{2}-\tilde\Delta\hat  E_j\right)+g^2\sum_j \hat E^2_j .
\end{align}

 
 \textit{Rydberg quantum simulator.--- } Our quantum simulator consists of a quasi-1D lattice of neutral atoms coupled to a Rydberg state $n\rm S$, with $n\gg 1$, by an effective Rabi frequency $\Omega$.
The atoms are initially trapped into a tweezers array~\cite{Barredo2018, PhysRevX.8.041054,PhysRevX.8.041055} and then released. 
By locally modulating the laser detuning, a configuration in which some atoms are excited to a Rydberg state and the others are in their internal ground state is created.
A non trivial dynamics is then induced by remodulating the laser detuning: 
The atoms move from ground to Rydberg states and thus interact among each other.
In the following we show that this process effectively reproduces the gauge invariant dynamics of the electric field in the $\mathbf{Z}_3$ model.
We start by showing that the gauge invariant electric field eigenstates are mapped into a set of atomic configurations in which the atoms are in their ground or Rydberg states. 

In general, two atoms at distance $r$ can be simultaneously excited only if $\Omega > V_r=c_6/r^6$, where $V_r$ is the Van der Waals interaction energy, due to the so-called Rydberg blockade mechanism.
$c_6$ depends on the atomic species and on the specific excited state~\cite{PhysRevLett.112.183002}.
We impose gauge invariance  by mapping gauge-breaking states into atomic configurations forbidden by the Rydberg blockade.

The  lattice is shaped as a prism with an equilateral triangular section, as shown in Fig.~\ref{fig:shape} (a). 
Sets of three atoms, called links, lay in planes perpendicular to the main prism axes. 
The links are placed at a fixed distance $d$ among each others while the distance between atoms in the same set is $\ell$. 
Their dynamics is described by the Hamiltonian 
\cite{Bernien2017,Schau2012,Labuhn2016}:
\begin{align}\label{eq:ryd_ham}
\hat H_{r}=\sum_{j,\vartheta}\left[\Omega\,
\hat \sigma_{j\vartheta}^x
-\Delta_{j}^{\vartheta}\hat n_{j}^{\vartheta} + \frac{1}{2}
\sum_{ \vartheta'\!,\,k}V_{j,k}^{\vartheta,\vartheta'}\hat n_{j}^{\vartheta} \hat n_{k}^{\vartheta'}\right],
\end{align}
where each atom is labeled by the link-index $j$ and $\vartheta$ which indicates the position inside the link (Fig.~\ref{fig:shape} (b)). 
The ground state of each atom $\ket{g}_{j\vartheta}$ is coupled to the excited state $\ket{r}_{j\vartheta}$ by the operator $\sigma_{j\vartheta}^x$ with Rabi frequency $\Omega$. 
The projector $\hat n_{j\vartheta}=\proj{r}_{j\vartheta}$ is multiplied by a detuning term with $\Delta_{j}^{\vartheta}=\Delta-\delta_{j\vartheta}>0$ and $\delta_{j\vartheta}\ll \Delta$. 
The interaction depends only on the distance between the atoms.
We call $V_\ell=V_{j,j}^{\vartheta,\vartheta'}$, $V=V_{j,j+1}^{\vartheta,\vartheta'}$ with $\vartheta\neq\vartheta'$ and $V_d=V_{j,j+1}^{\vartheta,\vartheta}$. 
We   neglect $V_{j,k}^{\vartheta,\vartheta'}$ with $|k-j|>1$ since they are much smaller than the other energies involved.
    \begin{figure}
    \centering
    \begin{tabular}{c}
     \includegraphics[width=7cm]{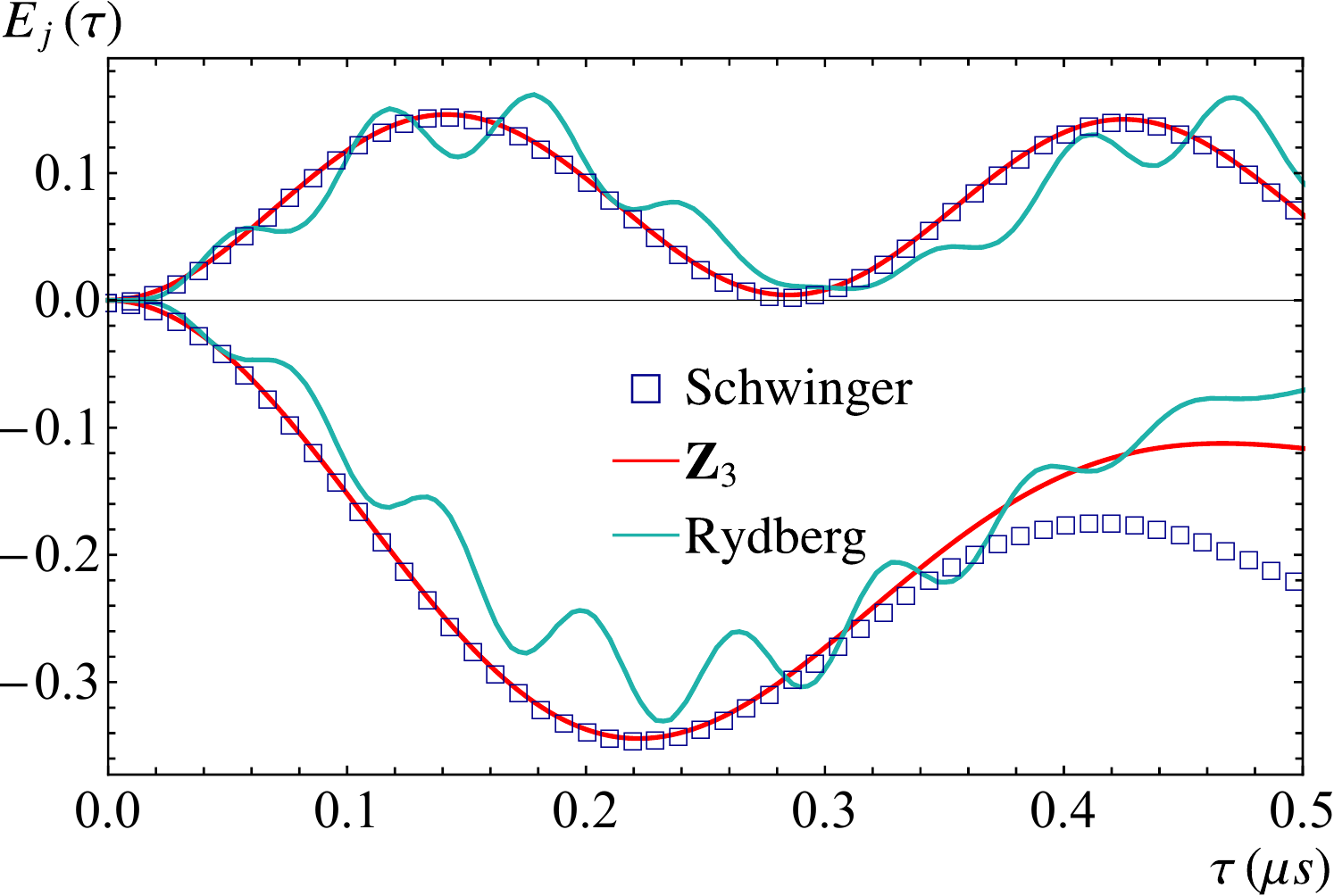}
     \end{tabular}
 \caption{\label{fig:vac}Electric field bulk local dynamics. 
 The dynamics of the quantum simulator (light green curve) and the $\mathbf{Z}_3$ model (red curve) exhibit a fair agreement both in for $\Gamma=4.0$ (upper curves, $j=10$) and $\Gamma=0.5$ (lower curves, $j=11$). 
 In the first case they coincide also with the dynamics of the Schwinger model (blue squares), as expected in the limit $\Gamma\gg 1$. 
 In the latter they deviate due to the truncation of the electric field spectrum in the $\mathbf{Z}_3$ model. 
 ($L=21$, $t=0.682\,\mathrm{MHz}$, $m=0$)}
  \end{figure} 

We first map the gauge invariant states into a set of atomic configurations and then we map the Hamiltonian $\hat H$ into  $\hat H_{r}$.
We  choose $\Delta \gg \Omega$ and $ V_\ell \gg \Omega$: for each single link, 
Rydberg excitations are enhanced by the large detuning but simultaneous excitations are prevented by Rydberg blockade.
 By applying second-order perturbation theory we restrict the dynamics of the $j$-th link to the subspace spanned by set of states $\Sigma_j=\{\ket{\vartheta}_j\}_{0\leq\vartheta\leq2}$ in which the atom in  position $\vartheta$ is excited (Fig.~\ref{fig:intro} (a)). 

We now consider a chain of L links, arranged as shown in Fig.~\ref{fig:shape},  with the distance $d$ between consecutive links such that $V_d\gg \Omega$ and $V\sim \Omega$: 
Simultaneous excitations of aligned atoms (red filled circles) are forbidden,  while non aligned excitations (blue filled circles) are allowed. 
We map  in a staggered fashion the electric field eigenstates $\ket{E}_j$ into the states $\ket{\vartheta}_j$ according to $\vartheta_j=(-E_j+4+(-1)^j)\mod 3$;
The set of allowed atomic configurations $\Phi \subset \bigotimes_{j=1}^L \Sigma_j$ corresponds to the set of the $\mathbf{Z}_3$ gauge invariant states.
In Fig.~\ref{fig:shape}~(a) we show two neighboring links and the site in between with charge $q_j=0, +1$: 
Since the electric field cannot decrease from the link $j-1$ to the link $j$, electric field states such as $\ket{0}_{j-1}\ket{-1}_{j}$ are mapped into configurations of excited atoms prevented by Rydberg blockade. 
On the other side, the configurations corresponding to the states $\ket{0}_{j-1}\ket{0}_{j}$ ($q_j=0$) or $\ket{0}_{j-1}\ket{1}_{j}$ ($q_j=1$) are allowed.

In order to map the Hamiltonian  $\hat H$ (\eq{eq:Z3_ham_g}) into $\hat H_r$ defined in \eq{eq:ryd_ham} we must confine the dynamics into the subspace spanned by $\Phi$:
By applying second order perturbation theory we obtain the effective Hamiltonian $\hat H_r^{\Phi}=-t\sideset{}{'}\sum_{\substack{j,\vartheta\neq\vartheta'}} \ket{\vartheta}\!\!\bra{\vartheta'}_j + \sum_{j, \vartheta}\delta_{j,\vartheta} \ket{\vartheta}\!\!\bra{\vartheta}_j $
where $t=\Omega^2(1/(\Delta-2V)+1/(V_l+2V-\Delta))$ and the energy shift   $L\Delta-(L-1)V$ has been applied (see Appendix \ref{app:Q_sim}). The primed sum is restricted to the transitions between states in $\mathrm{span}(\Phi)$ and is equivalent to the hopping term of the Hamiltonian $\hat H$ by construction.
The electric field energy coupling $g^2$ is obtained by modulating the local detuning $\delta_{j,\vartheta}$ such that $g^2=\delta_{j,\vartheta'}-\delta_{j,\vartheta}$, with $\ket{\vartheta'}\equiv \ket{E=\pm1}$ and $\ket{\vartheta}\equiv \ket{E=0}$.

The mass term involves an interaction between nearest-neighbor links in the Hamiltonian $\hat H$: 
Its implementation should be encoded in the inter-links interaction term of the Hamiltonian $\hat H_r$. 
The Hamiltonian $\hat H_r^{\Phi}$ implements the case with $m=0$:
Indeed, two-link states with or without charge are both represented by configurations whose interaction energy is $V$.
In Appendix~\ref{app:mass} we show that the case $m\neq 0$ can be implemented by modifying the geometry of the lattice.

  \textit{$\mathbf{Z}_3$ model dynamics.--- } 
  {The agreement between the dynamics of the $\mathbf{Z}_3$ and the Schwinger models depends on the parameter $\Gamma = g^2/t$.
  In the limit $\Gamma>1$ the $\mathbf{Z}_3$ model better approximates the Schwinger one: 
  Its dynamics is naturally constrained in the low energy sector due to the large electric field coupling and it is not affected by the truncation of the electric field spectrum.
As an example, we compare the dynamics of the two models by taking the bare vacuum, with $E_j=0\, \forall j$, as initial state: In Fig.~\ref{fig:vac} (upper panel) we show the local bulk dynamics of the electric field for $\Gamma=4$: 
The $\mathbf{Z}_3$ dynamics (red curve) and the Schwinger one (squares) coincide.
On the opposite regime, with $\Gamma=0.5$ the dynamics of the two models relaxes to different values (bottom panel of Fig.~\ref{fig:vac}).
  }
  
\begin{figure}
    \centering
    \begin{tabular}{cc}
     \includegraphics[width=4.cm]{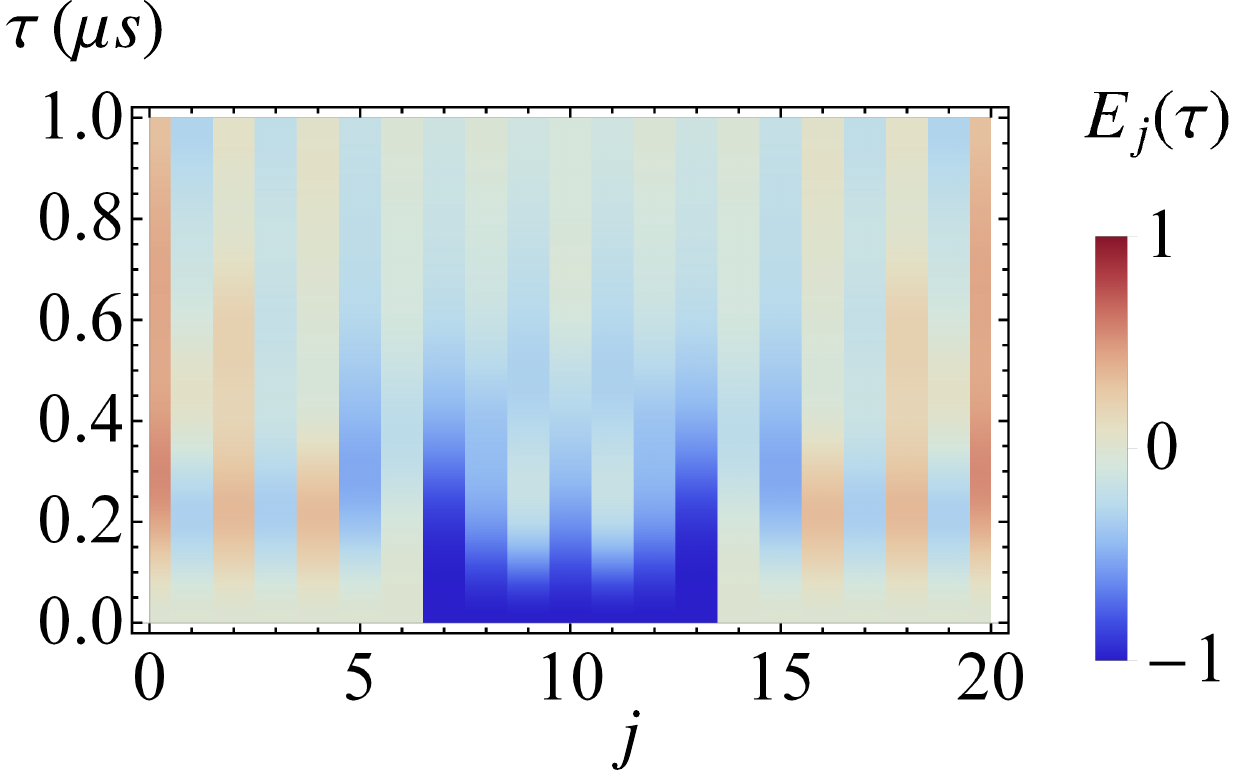} 
     & \includegraphics[width=4.cm]{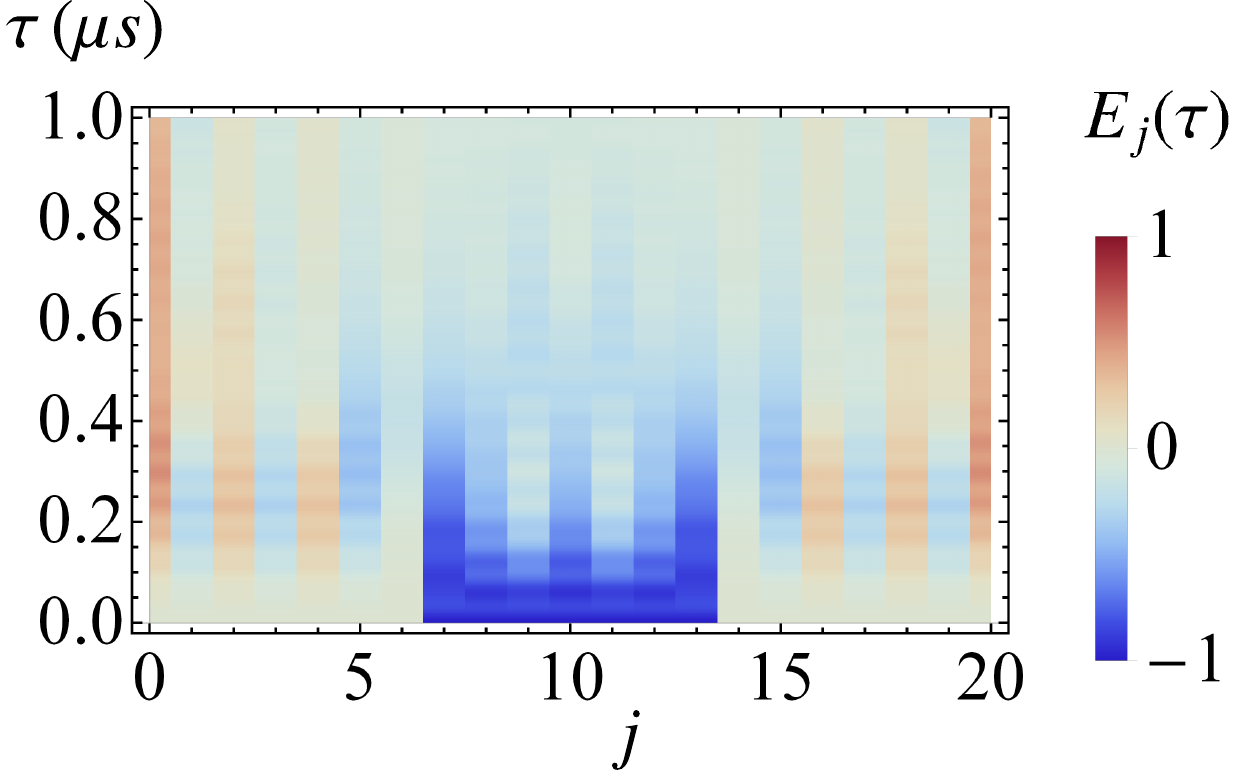} \\
     \includegraphics[width=4.cm]{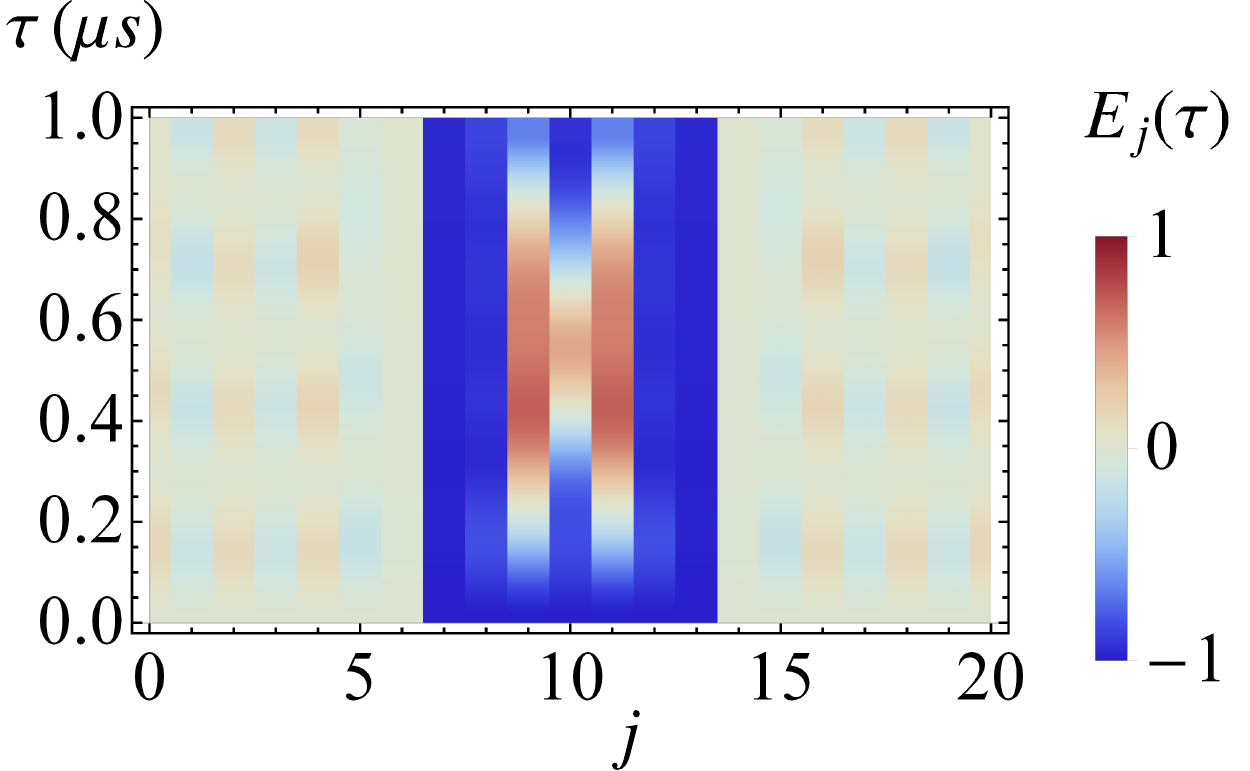}  
     & \includegraphics[width=4.cm]{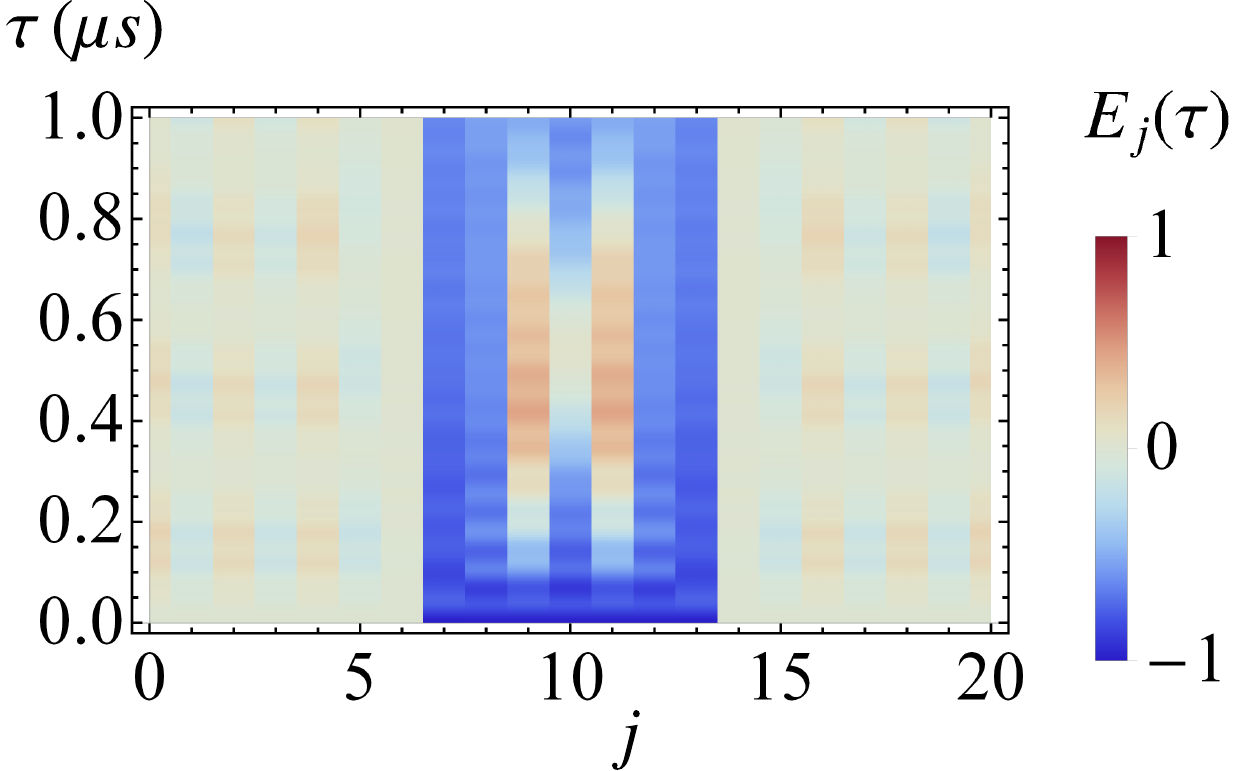}
     \end{tabular}
          \pgfputat{\pgfxy(-6.8,2.7)}{\pgfbox[left,top]{\footnotesize (a)}} 
           \pgfputat{\pgfxy(-2.6,2.7)}{\pgfbox[left,top]{\footnotesize (b)}} \\
          \pgfputat{\pgfxy(1.8,2.65)}{\pgfbox[left,top]{\footnotesize (c)}}
          \pgfputat{\pgfxy(6,2.95)}{\pgfbox[left,top]{\footnotesize (d)}}
 \caption{\label{fig:conf}(a),(c): String dynamics of the $\mathbf{Z}_3$ model. (b),(d): Rydberg-atom simulator dynamics. For $\Gamma=0.5$ (upper line) the initial string breaks;
for  $\Gamma=4.0$ (bottom line) the string is not broken during time evolution. 
The oscillation of the electric field expectation value inside the string is not due to string breaking but to the intrinsic dynamics of the $\mathbf{Z}_3$ model. ($L=21$, $t=0.682\,\mathrm{MHz}$, $m=0$)
 }
  \end{figure}

 {
 In order to investigate the string dynamics, we take as initial state $\ket{s(0)}$ an  electric field  string of length $s$ originated by a positron-electron pair: 
The state evolves as $\ket{s(\tau)}=\exp \{-i\hat H \tau\}\ket{s(0)}$ ($\hbar=1$)~\footnote{
The dynamics of both $\mathbf{Z}_{3}$ and $U(1)$ models has been obtained by exact diagonalization.}.
  Although the ground state of the Hamiltonian $\hat H$ is always confined, the unitary dynamics induced by quenching $\Gamma$ allows us to observe  different regimes~\cite{PhysRevX.6.011023}. 
For $\Gamma\lesssim 1$ fluctuations of local charge density lead to the creation of positron-electron pairs which annihilate the electric field between them 
and break the string:
We show an example of this dynamics in Fig.~\ref{fig:conf}~(a), where a chain of $L=21$ links has been prepared 
with an electron-positron string of length $s=7$.
By quenching $\Gamma$ from $\Gamma=0$ to $\Gamma=0.5$ the string breaks during its evolution. 
A different scenario emerges by choosing instead $\Gamma >1$: 
In Fig.~\ref{fig:conf}~(c) we take the same initial state and evolve it with $\Gamma=4.0$.
Due to the large gap between the zero and nonzero electric field states  the string breaking process is strongly off-resonant and does not occur {in accessible times}.
It is worth noting the oscillations in the middle of the string,which are a peculiar of the $\mathbf{Z}_3$ model:
The electric field is oscillating between the two local degenerate states $\ket{E=\pm 1}_j$ and the transient in which $\langle \hat E_j \rangle =0$ is due to their superposition during the  population inversion process.
 }

\textit{Results.--- } 
We benchmark the dynamics  of the quantum simulator via a numerical analysis. 
The experimental parameters we use refer to {$^{87}\mathrm{Rb}$ atoms} excited to the state $\ket{68\,\mathrm{S};m=1/2}$,  with $c_6=612 \ghz$.
We set $\Omega=3\mhz$, $\Delta=27\mhz$, $\ell=4\,\mu m$, $d=5.5\,\mu m$ so we have $t=0.682\mhz$. 
We set $\delta_{j,\vartheta}=0$   for $\vartheta,\,j$ such that $E_j=0$ and $\delta_{j,\vartheta}=g^2$   for $\vartheta,\,j$ such that $E_j=\pm1$.
By calling $V^{(2)}$ the amplitude  of next-nearest neighbor  interactions, we have $V^{(2)}\ll V,\,V_\ell,\,V_d$ by construction.
The values we  choose for the parameters of the Hamiltonian $\hat H_r$ guarantee that $V^{(2)}<t,g^2$.
%
%
We have numerically checked that, under these conditions, the dynamics of the $\mathbf{Z}_3$ model is not significantly affected by next-nearest neighbor interactions and therefore we neglect them~\footnote{
Next-nearest neighbors interactions between excited atoms add an extra term in the Hamiltonian $\hat H_r^\Phi$ which does not break the gauge invariance, being diagonal in the electric field configurations basis. We have numerically checked that the dynamics generated by the Hamiltonians $\hat H_r^\Phi$, ore equivalently $\hat H$, is not significantly affected by this term.}.
We use a TEBD algorithm~\cite{PhysRevLett.93.040502,SCHOLLWOCK201196}
to simulate the dynamics of the Rydberg Hamiltonian $\hat H_r$ (see Appendix~\ref{app:TEBD}) and compare it with the exact diagonalization of the  $\mathbf{Z}_3$ model for a chain of $L=21$ links. 
The implementation thus requires $60$ atoms and is achievable on the basis of a recent experiment~\cite{Bernien2017}. 
In Fig.~\ref{fig:vac} we compare the dynamics of the local electric field obtained from the Rydberg quantum simulator
against the exact $\mathbf{Z}_{3}$ one:
%
The curves relative to the  $\mathbf{Z}_3$ model (red line) and to the quantum simulator (light blue line) are in a fair agreement in both  the cases with $\Gamma > 1$ (upper panel) and $\Gamma< 1$ (lower panel).
The high frequency oscillations in the curve relative to the Rydberg dynamics are reminiscent of the second order processes in $\Omega$ which generate in the transitions between different states $\ket{\vartheta}_j$.

This Rydberg-atom quantum simulator is able to catch the string dynamics:
As we show in Fig.~\ref{fig:conf}~(b), (d), it is possible to distinguish the string breaking and persisting regimes predicted by the $\mathbf{Z}_3$ model.

 \textit{Conclusions and outlook.---}
In this work we have introduced a quantum simulator for the study of the real time dynamics of an Abelian quantum lattice gauge theory with scalable sizes of the lattice up to $\sim\!\!20$ sites. 
Thanks to the Rydberg blockade mechanism, which is able to guarantee an high reliability between the original model and the experimental realization in the case of local interactions, we explored the dynamics of the $\mathbf{Z}_3$ gauge model in different parameters regimes and with different initial states.
{We compared the dynamics of the  $\mathbf{Z}_3$ model with that of the $U(1)$ gauge theory, 
finding remarkable similarities of the string dynamics.}
Our quantum simulator is therefore a versatile and reliable experimental setup useful for investigating exotic properties of discrete and continuous Abelian lattice gauge theories~\cite{PhysRevLett.122.050403,PhysRevD.99.014503,magnifico2019mathbb}. 
Possible outlook for this work is the extension to two-dimensional theories, in continuity with  a recent proposal about the study of 2D pure gauge systems~\cite{celi2019emerging}, as well as the application of this protocol for simulating clock variables to different models such as time crystals~\cite{fan2019discrete, PhysRevB.99.104303}. 

\acknowledgements{We acknowledge insightful discussions with P. Zoller, M. Lukin, A. Celi, P. Silvi, S. Pascazio. 
This work is partially supported by the Italian PRIN 2017, the Horizon 2020 research and innovation programme under grant agreement No 817482 (PASQuanS), the QuantERA project QTFLAG and QuantHEP, the DFG via the TWITTER project. 
SN would like to thank the Erwin Schrödinger International Institute (ESI) in Wien for 
the hospitality and support during the Programme on “Quantum Simulation - from Theory to Applications”. 
SM would like to thank the Kavli Institute for Theoretical Physics (KITP) in Santa Barbara 
for the support and hospitality during the QSIM19 program. 
The authors acknowledge support by the state of Baden-Württemberg through bwHPC and the German Research Foundation (DFG) 
through grant no INST 40/467-1 FUGG (JUSTUS cluster). 
The authors acknowledge the CINECA award under the ISCRA initiative, for the availability 
of high performance computing resources and support.
}

\appendix

\section{Integrating the matter field}\label{app:Ham}

In this section we derive the Hamiltonian $ \hat H_S^g$ of Eq. (2) starting from $\hat H_S$ defined in  Eq. (1) of the main text.
The subspace we are considering contains those states $\ket{\Phi}$ such that  $ \hat Q_j\ket{\Phi}=0\,\quad \forall j$. 
 The basis vectors of this subspace, $ \{\ket{m,\mathcal{\mathcal{ E}}}\}$, are the eigenstates of a projector operator $\hat P$ which
 can be written as a product of local terms $\hat P = \prod_{j=1}^L \hat P_{j}$: 
 We express $\hat P_{j}$ in terms of the local computational basis basis $\prod_{j=1}^L\{ |m\rangle_j\}\otimes \{|\mathcal{\mathcal{\hat E}}\rangle_j\}$. 
 In particular, since the Gauss' law establishes a relation between the expectation value of the charge density operator $\hat n_j$ and the related electric field flux  $\Delta \mathcal{\mathcal{\hat E}}_j=\mathcal{\mathcal{\hat E}}_{j}-\mathcal{\mathcal{\hat E}}_{j-1}$, 
we define the following projectors:
\begin{equation}\label{eq:Pdedef}
\hat P_{\Delta \mathcal{E}_j}(n_{j})=\sum_\mathcal{E}\proj{\mathcal{E}}_{j-1}\otimes\proj{\mathcal{E}+\Delta \mathcal{E}^{n_j}}_{j}\,,
\end{equation}
with $
\Delta \mathcal{E}^{\,n_j}\equiv \mathcal{E}_{j}-\mathcal{E}_{j-1}=\frac{1-p_j}{2}-n_{j}\,.
$
With this representation we have that
\begin{equation}\label{app:Pj}
\hat P_{j}= \proj{0}_j\otimes\hat P_{\Delta \mathcal{E}_j}(0) + \proj{1}_j\otimes\hat P_{\Delta \mathcal{E}_j}(1) \,,
\end{equation}
namely $\hat P_{j}$ selects those configurations such  that the difference between the electric field values in neighboring links and the occupation of the site between them satisfy the Gauss' law. 

Now we  focus on the aforementioned Hamiltonian $\hat H_S^g$. 
We show that, if we consider two basis vectors   $\ket{m, \mathcal{E}}$ and  $\ket{m',\mathcal{E}'}$, it exists an operator $\hat H_S^g$ such that:
\begin{align}\label{eq:gauge_tr_ham}
& \bra{m,\mathcal{E}}\hat H_{S}\ket{m',\mathcal{E}'}
=\bra{\mathcal{E}} 
\hat H_{S}^g
\ket{\mathcal{E}'}.
\end{align}
In the following we explicitly computes the operator $\hat H_{S}^g$.
We start by considering the mass term at the site $j$:
\begin{align}
\bra{m, \mathcal{E}}\,\hat  \psi_j^\dagger\hat  \psi_j
 \otimes \mathbf{1}^\mathcal{E}
\, \ket{m',  \mathcal{ E}'}
=
\bra{\mathcal{E}} \, 
\left ( \frac{1-\hat p_j}{2}-\Delta \mathcal{\hat E}_j\right)
\,\ket{\mathcal{ E}'},
\end{align}
where $\mathbf{1}^\mathcal{E} $ is identity operator acting on the electric field space,
and we used the fact that all the basis vectors $\{\ket{m,\mathcal{E}}\}$ satisfy the Gauss' law.
The computation of the electric field energy is straightforward since
 it is diagonal in the electric field local computational basis. 
 %

The hopping term is composed by the unitary propagator, which provides the evolution of the gauge field,
and the fermionic operators, which constraint  the allowed transitions. 
In order to integrate out the matter field, we need to recast the fermionic constraints in terms of the gauge field operators only. 
To this purpose we observe that
\begin{align}
&\,\bra{m,  \mathcal{E}}\, \hat \psi_j^\dagger\, \mathcal{\hat U}_j^\dagger\,\hat  \psi_{j+1}\,  \ket{m',  \mathcal{E}'}\\  \nonumber
=&\,\bra{m,  \mathcal{E}}\,\hat P\, \hat \psi_j^\dagger\, \mathcal{\hat U}_j^\dagger\,\hat  \psi_{j+1}\, \hat P\,  \ket{m',  \mathcal{E}'} \\ \nonumber
=&\,\bra{m,  \mathcal{E}}\, \hat P\,  \hat P_{j}\otimes   \hat P_{j+1} \,\hat  \psi_j^\dagger\,\mathcal{\hat U}_j^\dagger\, \hat \psi_{j+1}\,   \hat P_{j}\otimes   \hat P_{{j+1}}\, \hat P \,  \ket{m',  \mathcal{E}'}\,,
\end{align}
where we have used the definition of $\hat P$ 
as well as the fact that $\hat P_{j}^2=\hat P_{j} \, \forall \, j$.
Discarding the overall projector $\hat P$ and using the definition (\ref{app:Pj}) we obtain
\begin{align}\label{app:conto}
&\,\bra{m,  \mathcal{E}}\, \left[ \proj{1}_j\otimes\hat P_{\Delta ,  \mathcal{E}_j}(1)\right] \\ \nonumber
&\otimes \left[ \proj{0}_{j+1}\otimes\hat P_{\Delta \mathcal{E}_{j+1}}(0)\right] \,\hat  \psi_j^\dagger\,\mathcal{\hat U}_j^\dagger\, \hat \psi_{j+1} \\   \nonumber
&\left[\proj{0}_{j}\otimes\hat P_{\Delta \mathcal{E}_{j}}(0)\right]\otimes\left[\proj{1}_{j+1}\otimes\hat P_{\Delta \mathcal{E}_{j+1}}(1)\right]  \,  \ket{m',  \mathcal{E}'}\\  \nonumber
=&\,\bra{m,  \mathcal{E}}\,   \left[\hat P_{\Delta  \mathcal{E}_j}(1)  \hat P_{\Delta  \mathcal{E}_{j+1}}(0) \,\mathcal{\hat U}_j^\dagger\, \hat P_{\Delta \mathcal{E}_{j}}(0) \hat P_{\Delta \mathcal{E}_{j+1}}(1)\right] \\ \nonumber
&\otimes  |1\rangle\langle0|_{j}\otimes|0\rangle\langle1|_{j+1}\,  \ket{m',  \mathcal{E}'}\\ \nonumber
= &\, 
\bra{\mathcal{E}} \, 
\left[\hat P_{\Delta   \mathcal{E}_j}(1)  \hat P_{\Delta   \mathcal{E}_{j+1}}(0) \,\mathcal{\hat U}_j^\dagger\, \hat P_{\Delta \mathcal{E}_{j}}(0) \hat P_{\Delta  \mathcal{E}_{j+1}}(1)\right]
 \, \ket{\mathcal{E}'} \\  \nonumber
=&\, \bra{ \mathcal{E}}\, 
\left[\mathcal{\hat U}_j^\dagger\, \hat P_{\Delta  \mathcal{E}_{j}}(0) \hat P_{\Delta   \mathcal{E}_{j+1}}(1) \right] \, 
\ket{ \mathcal{E}'}.
\end{align}
The first passage is justified by observing that $\hat \psi_j^\dagger \hat \psi_{j+1}=|1\rangle\langle0|_{j}\otimes|0\rangle\langle1|_{j+1}$.
\footnote{This is true since the product of the  nearest-neighbor operators $\hat \psi_j^\dagger$and $ \hat \psi_{j+1}$ is bosonic.}
The second passage is a consequence of the definition (\ref{eq:Pdedef}) which allows to write
\begin{align}
&\bra{m}[|1\rangle\langle0  |_{j} \otimes |0\rangle\langle1|_{j+1} ]\ket{m'} =1\qquad \Longleftrightarrow   \\ 
&\bra{ \mathcal{E}} \left[\hat P_{\Delta   \mathcal{E}_j}(1)  \hat P_{\Delta   \mathcal{E}_{j+1}}(0) \,\mathcal{\hat U}_j^\dagger\, \hat P_{\Delta  \mathcal{E}_{j}}(0) \hat P_{\Delta  \mathcal{E}_{j+1}}(1)\right] \ket{  \mathcal{E}'} =1 \nonumber.
\end{align}
In the last passage of \eq{app:conto}, since
$
\hat P_{\Delta \mathcal{E}_j}(1)  \hat P_{\Delta \mathcal{E}_{j+1}}(0)=\mathcal{\hat U}_j^\dagger\, \hat P_{\Delta \mathcal{E}_{j}}(0) \hat P_{\Delta \mathcal{E}_{j+1}}(1)\mathcal{\hat U}_j
$,
it follows
\begin{align}
&\,\hat P_{\Delta \mathcal{E}_j}(1)  \hat P_{\Delta \mathcal{E}_{j+1}}(0)\,\mathcal{\hat U}_j^\dagger\, \hat P_{\Delta \mathcal{E}_{j}}(0) \hat P_{\Delta \mathcal{E}_{j+1}}(1) \\ \nonumber
=&\, \mathcal{\hat U}_j^\dagger\, \hat P_{\Delta \mathcal{E}_{j}}(0) \hat P_{\Delta \mathcal{E}_{j+1}}(1)\mathcal{\hat U}_j\, \mathcal{\hat U}_j^\dagger\, \hat P_{\Delta \mathcal{E}_{j}}(0) \hat P_{\Delta \mathcal{E}_{j+1}}(1) \\ \nonumber
=&\,  \mathcal{\hat U}_j^\dagger\, \left(\hat P_{\Delta \mathcal{E}_{j}}(0) \hat P_{\Delta \mathcal{E}_{j+1}}(1)\right)^2 
= \mathcal{\hat U}_j^\dagger\, \hat P_{\Delta \mathcal{E}_{j}}(0) \hat P_{\Delta \mathcal{E}_{j+1}}(1).
\end{align}
As a conclusion, we have derived the Hamiltonian 
\begin{align}
\hat H_{S}^g =&\, -t\sum_{j}\mathcal{\hat  U}_j^\dagger\, \hat P_{\Delta \mathcal{E}_{j}}(0) \hat P_{\Delta \mathcal{ E}_{j+1}}(1) + \mathrm{h.c.}  \\ \nonumber
&\,+ m \sum_j \left( \frac{1-p_j}{2}-\Delta\mathcal{\hat E}_j\right)+g^2\sum_j \mathcal{\hat E}^2_j \,.
\end{align}


\section{Second-order derivation of the Hamiltonian $\hat H_r^\Phi$}\label{app:Q_sim}
In this section we show how to map the Rydberg Hamiltonian to the $\mathbf{Z}_{3}$ 
Hamiltonian (3) by using second-order perturbation thoery.
We start by considering the single link case and then we extend the result to the entire chain.
The Hamiltonian $\hat H_r$ (Eq. (4) of the main text), 
in terms of single link $3$-atom states
$
\{
|rgg\rangle,\, 
|grg\rangle ,\, 
|ggr\rangle ,\, 	
|ggg\rangle ,\, 
|grr\rangle ,\, 
|rgr\rangle ,\, 	
|rrg\rangle ,\, 
|rrr\rangle 	
\}
$
reduces to the following $8\times 8$ matrix
\begin{widetext}
\begin{equation} \label{eq:ham_matr}
\hat H_{r,s} = \left(
\begin{array}{ccc||c|ccc|c}
\delta_0 & 0 & 0 & \Omega & 0 & \Omega & \Omega & 0\\
 0 & \delta_1 & 0 & \Omega & \Omega & 0 & \Omega & 0\\
0 & 0 & \delta_2 & \Omega & \Omega & \Omega & 0 & 0\\
\hline\hline
\Omega & \Omega & \Omega & \Delta & 0 & 0 & 0 & 0\\
\hline
0          & \Omega & \Omega & 0 & V_\ell-\Delta & 0 & 0 & \Omega \\
\Omega & 0           & \Omega & 0 & 0 & V_\ell-\Delta & 0 & \Omega \\
\Omega & \Omega & 0 & 0 & 0 & 0 & V_\ell-\Delta & \Omega \\
\hline
0 & 0 & 0 & 0 & \Omega & \Omega & \Omega & 2(V_\ell-\Delta)
\end{array}
\right)
\equiv
 \left(
\begin{array}{c||c}
&\\
\hat H^{PP}_{r,s} & \hat H^{PN}_{r,s} \\
&\\
\hline\hline
&\\
\hat H^{NP}_{r,s} & \hat H^{NN}_{r,s}\\
&
\end{array}
\right). 
\end{equation}
\end{widetext}
The first three states contain only one excited atom and they form the set $\Sigma=\{\ket{\vartheta}\}_{0\leq\vartheta\leq 2}$ defined in the main text (links indexes are omitted here). 
We have applied a shift of the energy equal to $\Delta$ so the energy of the states in $\Sigma$ is $E_s\sim \delta_\vartheta$.
We call $\mathcal{H}_s^P$ the subspace spanned by $\Sigma$ and $\mathcal{H}_s^N$ the complementary one such that $\mathcal{H}_s=\mathcal{H}_s^N\oplus\mathcal{H}_s^P$.
We defined the projectors $\hat P_s$ and $\hat N_s=\mathbf{1}-\hat P_s$ such that 
\begin{align}
\hat H_{r,s}^{PP} &=\hat P_{s} \hat H_{r,s} \hat P_{s}  \\ \nonumber
\hat H_{r,s}^{NN} &=\hat N_{s}  \hat H_{r,s} \hat N_{s}  \\ \nonumber
\hat H_{r,s}^{PN} &=\hat P_{s}  \hat H_{r,s} \hat N_{s}  = (\hat H_{r,s}^{NP})^\dagger.
\end{align}
They correspond respectively to the top left, bottom right and off-diagonal parts of $\hat H_{r,s}$ delimited by double lines in Eq. (\ref{eq:ham_matr}).
The effective Hamiltonian relative to the subspace $\mathcal{H}^{P}_{s}$ can be derived by assuming that there exists a set of eigenstates of $\hat H_{r,s}$ whose energies are perturbations of the spectrum of $\hat H_{r,s}^{PP}$. Let us consider  an eigenstate $|\Psi\rangle$ whose energy satisfies 
${E_s\sim \delta_\vartheta\ll V_{\ell}-\Delta,\Delta}$ 
and define $|\Psi_P\rangle = \hat P_{s} |\Psi\rangle$ and $|\Psi_N\rangle = \hat N_{s} |\Psi\rangle$; 
the eigenvalue equation can be written as

\begin{equation}\label{eq:eig_eq}
  E_s\left[\!
     \begin{array}{c}
       |\Psi_P\rangle \\
       |\Psi_N\rangle \\
     \end{array}
  \! \right]=\left[\!\begin{array}{cc}
                        \hat H_{r,s}^{PP} & \hat H_{r,s}^{PN} \\
                        \hat H_{r,s}^{NP} & \hat H_{r,s}^{NN} \\
                      \end{array}
                    \right]\left[\!
     \begin{array}{c}
       |\Psi_P\rangle \\
       |\Psi_N\rangle \\
     \end{array}
   \!\right],
\end{equation}
from which it emerges that $|\Psi_P\rangle$ obeys the equation
\begin{align} \nonumber
E_{s} |\Psi_P\rangle &= \left[ \hat H_{r,s}^{PP} + \hat H_{r,s}^{PN} \frac{1}{E_{s}- \hat H_{r,s}^{NN}} \hat H_{r,s}^{NP} \right] |\Psi_P\rangle \\
&= \hat{\tilde H}_{r,s}^{PP} |\Psi_P\rangle.
\end{align}

We compute the matrix elements of $ \hat{\tilde H}_{r,s}^{PP}$  in the basis of the states $\{ |rgg\rangle, |grg\rangle, |ggr\rangle \}$. 
We consider its action on the state $|rgg\rangle$:
\begin{align} \label{eq:Hpptilde}
&\hat{\tilde  H}_{r,s}^{PP} |rgg\rangle  \\ \nonumber
=\,& \delta_0 |rgg\rangle + \Omega \hat H_{r,s}^{PN} \frac{1}{E_{s}- \hat H_{r,s}^{NN}} (|ggg\rangle + |rrg\rangle +  |rgr\rangle).
\end{align}
We approximate the operator $\hat H_{r,s}^{NN}$ with its diagonal contribution,
since $\Omega$ is much smaller than $\Delta,V_\ell-\Delta$. In \eq{eq:Hpptilde}, we use
$
1/(E_s-\hat H_{r,s}^{NN})\simeq - (\hat H_{r,s}^{NN})_{diag}^{-1}
$,
where we have neglected $E_{s}\ll \Delta,V_\ell-\Delta$, and
therefore we get
\begin{align}
& \hat{\tilde H}_{r,s}^{PP}|rgg\rangle \\ \nonumber
&\simeq \delta_1 |rgg\rangle - \Omega \hat H_{r,s}^{PN}(\hat H_{r,s}^{NN})_{diag}^{-1} (|ggg\rangle + |rrg\rangle +  |rgr\rangle) \\ \nonumber
&=\delta_1 |rgg\rangle -  \Omega \hat H_{r,s}^{PN}\left[ \frac{1}{\Delta}|ggg\rangle +\frac{1}{V_\ell-\Delta}( |rrg\rangle +  |rgr\rangle)\right] \\ \nonumber
&=\left[\delta_1-\Omega^2\left(\frac{1}{\Delta}+\frac{2}{V_\ell-\Delta}\right)\right]|rgg\rangle \\ \nonumber
&+\Omega^2\left(\frac{1}{\Delta}+\frac{1}{V_\ell-\Delta}\right)|grg\rangle +\Omega^2\left(\frac{1}{\Delta}+\frac{1}{V_\ell-\Delta}\right)|ggr\rangle \,.
\end{align}
By repeating the same procedure for the states $|grg\rangle$ and $|ggr\rangle$ we obtain the following effective Hamiltonian
\begin{equation}
\hat H_{r,s}^\Phi=-t_s\sum_{\vartheta\neq\vartheta'}\ket{\vartheta}\!\!\bra{\vartheta'}+\sum_{\vartheta}\delta_\vartheta\proj{\vartheta}\,,
\end{equation}
with $t_s =\Omega^2\left( \frac{1}{V- \Delta}+\frac{1}{\Delta} \right) $, 
where we are neglecting the overall constant  $-\Omega^2\left(\frac{2}{V-\Delta}+\frac{1}{ \Delta} \right) $.

When we consider a chain of $L$ links we must take into accounts the inter-link energies $V$ and $V_d$.
%
 %
 In fact, the interactions between nearest-neighbour modify the local Hamiltonian $\hat H^{N N}_{r,s}$.
 Transitions between to different gauge invariant states $\ket{\Psi}$ and $\ket{\Psi'}$
 are mediated by local link changes $\ket{\vartheta}_{j}\to\ket{\vartheta'}_{j}$.
 
 For example, let us suppose the transition $\ket{\Psi} \to \ket{\Psi'}$ corresponds
 to only changing the local state $\ket{rgg}_{j}$ to the the state $\ket{grg}_{j}$.
 This transition is now mediated by the doubly excited local state $\ket{rrg}_{j}$
 and by the local state $\ket{ggg}_{j}$ with no excitations.
 %
 Due to the inter-link interaction, 
 the transition $\ket{rgg}_{j}\to\ket{rrg}_{j}$
 induces a local change in the energy
 $V_\ell + 2 V-\Delta$.
%
%
Analogously, the transition $\ket{rgg}_{j}\to\ket{ggg}_{j}$
leads to the an energy change $\Delta-2 V$. 
As a consequence, the allowed transition $\ket{rgg}_{j}\to\ket{grg}_{j}$
is a second-order process with rate
${t=\Omega^2\left( \frac{1}{V_\ell+2 V- \Delta}+\frac{1}{\Delta-2V} \right)}$.

So far, we only consider transition between gauge invariant many-body states. 
However, the Hamiltonian $\hat H_{r}$ may allow local transitions such as 
${\ket{rgg}_{j}\ket{ggr}_{j+1}\to\ket{rgr}_{j}\ket{ggr}_{j+1} \to \ket{ggr}_{j}\ket{ggr}_{j+1} }$ 
with two excited atoms at distance $d$. 
Such transition is suppressed by an energy penalty $V_{d}$ and breaks the gauge invariance.
As far as $V_{d} \gg V,\, \Omega$, these second-order processes can be neglected.
In conclusion, we have shown how $\hat H_r$ reduces to
$\hat H_r^\Phi$ up to second-order corrections in $\Omega$.

\section{Mass-term implementation}\label{app:mass}

In this section we describe how to implement a non-zero mass term in the Hamiltonian $\hat H_{r}$.
Let us recall that, in the $\mathbf{Z}_{3}$ model, even matter sites can contain 
a zero or negative charge corresponding respectively to a zero or $m>0$ mass energy.
If we consider the electric-field Hamiltonian $\hat H$, given an even site $j$, 
the energies of the two local states ${|E\rangle_{j-1}\otimes|(E-1)\!\mod 3\rangle_{j}}$ and
$|E\rangle_{j-1}\otimes|E\rangle_{j}$ differ by $m$. Similar considerations apply for odd sites.

\begin{figure}
    \centering
 \includegraphics[height=3cm]{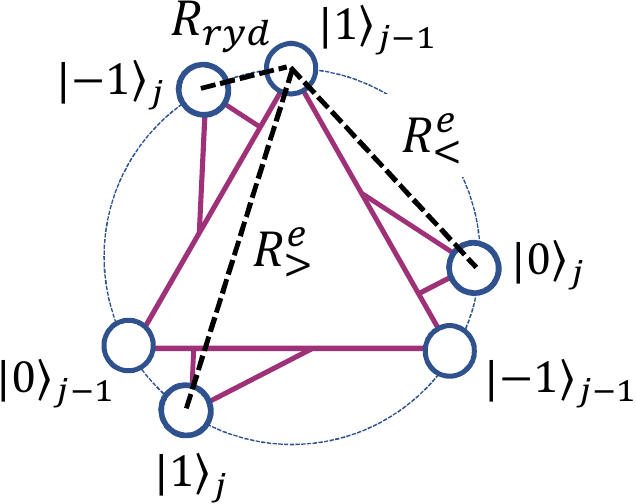}
    \caption{\label{fig:gauss1} Mass implementation: In the case with $m\neq0$ the positions of the atoms belonging to even links are rotated anticlockwise by an angle $\phi$. The correspondent value of the mass is $m=V_c^e-V_v^e$ is implemented.}
  \end{figure}

In order to achieve this condition, we tilt the triangular structures relative 
to the even links by a small angle $\phi$ in an anticlockwise way, 
as shown in Fig. \ref{fig:gauss1}.
A rotation in the plane perpendicular to the lattice axis 
makes the distances between the site corresponding to $E_{j-1}=-1$ and the sites $E_{j}=-1$ and $E_{j}=0$ to be different: 
In this way, different interaction strengths are engineered and thus the energy difference between the vacuum and the charged configuration can be implemented.
We define the following characteristic inter-link distances:
\begin{align}
R_{ryd}&= [d^2 +4\, \ell^2 \sin^2(\phi/2)]^{1/2}\,,\\
R_> &=  [d^2 +4/3\, \ell^2 \sin^2(\pi/3+\phi/2)]^{1/2}\,,\\
R_<&=  [d^2 +4/3\, \ell^2 \sin^2(\pi/3-\phi/2)]^{1/2}.
\end{align}
$R_{ryd}$ is the smallest distance, corresponding to forbidden configurations, 
while $R_{\lessgtr}$ correspond to the allowed ones.

The energies corresponding to charged and vacuum  
configurations are  $V_c^e=c_6/R_<^6$ and  $V_v^e=c_6/R_>^6$, respectively. 
Analogously, we define the same energies for odd matter sites, namely $V_c^o=c_6/R_<^6$ and  $V_v^o=c_6/R_>^6$.

Staggering is implemented by a further lattice deformation: 
Indeed, the energy of the vacuum configuration for an even matter site must be equal to the energy of 
the charged configuration for an odd matter site, i.e. $V_v^e=V_c^o$. 
In order to achieve the above statement 
we change the inter-links distance of the chain by a small amount  $\varepsilon$, such that $d^{o}=d+\varepsilon$ and $d^e=d-\varepsilon$ 
relative to odd and even sites respectively.
We choose a value of $\phi$ and thereafter choose a value of $\varepsilon$ to satisfy the condition $V_v^e=V_c^o$.
As a result we get $m=V_c^e-V_v^e\simeq V_c^o-V_v^o$. 
For example, with the parameters used in the main text, by applying a rotation $\phi=0.05\,\mathrm{rad}$ we obtain $t=0.667\mhz$, $m_{e^-}=V_c^e-V_v^e=0.385 \mhz$ and $m_{e^+}=V_c^o-V_v^o=0.356 \mhz$. 
We can simulate therefore a mass $m=(m_{e^+}+m_{e^-})/2$ with a relative error $\sim4\%$. 


\section{TEBD-MPS numerical simulation details}\label{app:TEBD}

The dynamics of the Rydberg atoms quantum simulator has been numerically simulated by using the Hamiltonian $\hat H_r$ defined in Eq. (4) of the main text. 
We considered only interactions between nearest-neighbour links. 
We used the MPS representation of the many-body state. 
We took the link as local subspace: 
Since it is composed by three atoms which can be in a Rydberg or in the internal ground state, 
its Hilbert space dimension is $2^3=8$. 
The auxiliary dimension was set to $128$ and 
we checked the convergence of the dynamics by repeating the same simulations with larger bond dimension equal to $256$. 
Note that the dynamics of each link is mostly constrained in the three-dimensional subspace spanned by $\Sigma_j$, 
allowing accurate results with a relatively small bond dimension.

The dynamics has been computed by using the TEBD algorithm with second-order Suzuki-Trotter decomposition of the evolution operator,
with time-step $\mathrm{dt}=0.01\, (2 \pi)^{-1}\, \mu s$.
Local and interaction parameters have been chose such that 
$\{\Delta=27\mhz,\,V_\ell=149.414\mhz,\,\Omega=3\mhz, \delta_\vartheta=g^2\}$
and
$\{V=6.186 \mhz,\,V_d=22.109\mhz\}$. 


%

\end{document}